# Material and device characterization of Type-II InAs/GaSb superlattice infrared detectors

*M. Delmas[1*], M. C. Debnath[2], B. L. Liang[2], D. L. Huffaker[1,2]*

[1]*School of Physics and Astronomy, Cardiff University, The Parade, Cardiff CF243AA, UK*

[2]*California NanoSystem Institute, University of California, Los Angeles, CA 90095, USA*

## Abstract

This work investigates midwave infrared Type-II InAs/GaSb superlattice (SL) grown by molecular beam epitaxy on GaSb substrate. In order to compensate the natural tensile strain of the InAs layers, two different shutter sequences have been explored during the growth. The first one consists of growing an intentional InSb layer at both interfaces (namely GaSb-on-InAs and InAs-on-GaSb interfaces) by migration enhanced epitaxy while the second uses the antimony-for-arsenic exchange to promote an 'InSb-like' interface at the GaSb-on-InAs interface. SLs obtained via both methods are compared in terms of structural, morphological and optical properties by means of high-resolution x-ray diffraction, atomic force microscopy and photoluminescence spectroscopy. By using the second method, we obtained a nearly strain-compensated SL on GaSb with a full width at half maximum of 56 arcsec for the first-order SL satellite peak. Relatively smooth surface has been achieved with a root mean square value of about 0.19 nm on a 2 µm x 2 µm scan area. Finally, a p-i-n device structure having a cut-off wavelength of 5.15 µm at 77K has been demonstrated with a dark-current level of 8.3 x $10^{-8}$ A/$cm^2$ at -50 mV and a residual carrier concentration of 9.7 x $10^{14}$ $cm^{-3}$, comparable to the state-of-the-art.

*Corresponding author: Tel: +44 2922510182. e-mail address: DelmasM@Cardiff.ac.uk

1. Introduction

High performance photodetectors operating in the midwave (3 - 5 µm) and longwave (8 - 12 µm) infrared (IR) spectral domain are useful for a broad range of civil, industrial and spatial applications. Since the proposition made by Smith and Mailhiot [1], the Type-II InAs/GaSb superlattice (SL) material has been of great interest as it offers unique properties for infrared detection including a high absorption coefficient, low dark-currents and the possibility to address a wide range of wavelengths by tuning the SL band edges [2].

The SL is usually grown by molecular beam epitaxy (MBE) which permits precise control of each layer thickness. One of the key challenges in SL growth is that the InAs binary has a smaller lattice parameter than GaSb ($\Delta a/a \sim -0.6\%$) and that the SL layer is therefore naturally under tensile strain on GaSb substrate. To achieve high quantum efficiency and low dark-current, it is crucial to develop growth methods that allow growing a thick SL layer with low defect and dislocation density. The control of both interfaces (IFs), namely the InAs-on-GaSb and GaSb-on-InAs IFs, is of significant importance to obtain a strain-compensated SL layer with high structural and optical quality. Depending on the shutter sequence during the growth, InSb bonds ('InSb-like' IF) or GaAs bonds ('GaAs-like' IF) can be formed. By introducing an 'InSb-like' IF, the tensile strain in the InAs layers can be compensated by the compressive strain in InSb ($\Delta a/a \sim +7.8\%$ with GaSb) resulting in an enhancement of the device performance [3] [4] [5].

Different growth methods have been reported to form an 'InSb-like' IF at the GaSb-on-InAs IF. One of them uses the antimony-for-arsenic exchange where each InAs layer of the SL are exposed to an antimony soak for few seconds [6] [7]. Another method consists of growing an intentional InSb layer either by conventional MBE (The indium and antimony shutters are synchronously opened) [5] [8] [9] or by migration enhanced epitaxy (MEE) (The indium and antimony shutters are asynchronously opened) [10] [11]. It has been demonstrated that abrupt interfaces can be grown by MEE [12] [13] and, in the case of SLs it has led to a smoother surface with improved optical properties compared to an InSb IF layer grown by conventional MBE [14].

In this paper, we compare two SL structures having a cut-off wavelength close to 5 µm at 77K. The first structure has been grown with an intentional InSb layer grown by MEE at both interfaces while for the second, an 'InSb-like' IF is formed using the antimony-for-arsenic exchange at the GaSb-on-InAs IF. Both growth methods are compared in terms of structural, morphological and optical properties by means of high-resolution x-ray diffraction (XRD), atomic force microscopy (AFM) and photoluminescence (PL) spectroscopy. Following these results, we present the electrical performances of a p-i-n photodiode grown using the appropriate shutter sequence for the SL layer.

## 2. Experimental details

All InAs/GaSb SL structures presented in this paper were grown on a quarter of two-inch p-type (001)-oriented GaSb substrate in a Veeco Gen 930 molecular beam epitaxy reactor equipped with dual filament SUMO Knudsen effusion cells for gallium (Ga) and indium (In) and MARK V valved cracker effusion cells for arsenic (As) and antimony (Sb). The In and Ga growth rates were set to 0.3 and 0.5 ML/s, respectively. The InAs and GaSb layers were grown using a V/III flux ratio which is calibrated from RHEED oscillations of 1.2 and 2, respectively. The growth temperature was monitored by pyrometer and thermocouple, calibrated with the (1 x 3) to (2 x 5) reconstruction transition on the GaSb substrate and buffer surface. Before the growth, the native oxide on the GaSb substrate was first thermally desorbed at 540°C. The growth temperature was then lowered down to 490°C for the buffer layer and to 410°C for the SL layer to maintain a good quality of the 'InSb-like' IF [8].

We studied the shutter sequence during the growth of two different test samples (sample A and B) which the structure consists of a 50 nm undoped GaSb buffer, followed by 100 pairs of undoped InAs/GaSb SL and a 4 monolayers (ML) thick undoped GaSb capping layer. The SL period is composed of 7 ML of InAs and 4 ML of GaSb and the targeted cut-off wavelength is around 5 µm at 77K [15] [16]. The shutter sequence used for the SL layer growth of sample A and B is schematically presented in Figure 1. It is worth noting that the Sb flux is kept constant during the sample growth. For sample A, an intentional InSb layer is grown by MEE at both IFs. After the growth of the InAs layer, only the In shutter remains open for 1 s. It is then closed and, only the Sb shutter is opened for 6 s to saturate the In surface with Sb. The Ga shutter is then opened to grow the GaSb layer. At the InAs-on-GaSb IF, the Sb shutter remains open for an additional 6 s and, then only the In shutter is opened for another 1 s before growing the InAs layer. For sample B, the InAs layer is exposed to an incident Sb flux for 6 s to promote the Sb-for-As exchange and the formation of InSb bonds at the GaSb-on-InAs IF. The Ga shutter is then opened to grow the GaSb layer. Next, both Ga and Sb shutters are simultaneously closed while the In and As shutters are opened to grow the InAs layer. Note that in this case, the InAs-on-GaSb IF is not controlled whereas some research groups are exposing the GaSb surface to an incident As flux to promote a 'GaAs-like' IF [6] [7] [8].

Following the growth, the structural and morphological quality of both samples were studied using a Bede D1 X-ray diffractometer and a XE-100 Park Systems atomic force microscope. To access the optical properties of the SL structures, samples were then loaded in a cryostat equipped with $CaF_2$ windows to carry out PL measurements. The samples were optically excited using a 785 nm laser diode modulated at a frequency of 20 kHz. To collect the signal, a Nicolet iS50R Fourier transform infrared spectrometer equipped with an MCT detector was used.

### 3. Material characterizations

From the XRD spectra of the ω/2θ scan around the GaSb (004) reflection represented in Figure 2, it can be seen that the SL layer of sample A is under compressive strain on GaSb with a lattice mismatch of about $\Delta a/a \sim 0.479\%$. This indicates that the total InSb layer within the period, which is in compressive strain on GaSb, is too thick to obtain a strain-compensated SL. However, a nearly strain-compensated ($\Delta a/a \sim 0\%$) SL layer was obtained for sample B. The measured SL period thickness is 3.69 and 3.54 nm for sample A and B, respectively. This difference is expected since an additional InSb layer is grown at both IFs in the SL layer of sample A. Finally, sample B presents a smaller full width at half maximum (FWHM) of 56 arcsec for the first-order SL satellite peak (SL$_{-1}$) compared to 73 arcsec for sample A suggesting that sample B has a better structural quality than sample A.

The surface morphology for both samples has then been inspected by AFM measurements for different scan area (2 µm x 2 µm, 4 µm x 4 µm and 10 µm x 10 µm). The root mean square (RMS) values extracted from these measurements are summarized in Table 1. The surface roughness seems to be slightly lower for sample B whatever the scan area indicating that sample B has a relatively smoother surface morphology than sample A.

Finally, PL measurements have been carried out at a temperature of 77K (Figure 3) using the experimental procedure described in section II. The wavelength at the maximum peak intensity is equal to 5.55 µm (223 meV) and 5.10 µm (243 meV) for sample A and B, respectively. This shift to longer wavelength (i.e. smaller energy) observed for sample A confirms that the total InSb layer within the period is thicker than sample B. Indeed, it is consistent with previous work where the PL peak shifts from 245 meV to 215 meV with the increase of the InSb IF layer thickness (from 0 ML to 1.5 ML) at the GaSb-on-InAs IF [8]. Additionally, both samples show similar PL intensity and a FWHM PL peak of about 26 meV.

From those results, it seems that even though we introduced a compressive strain in sample A, the material quality does not appear degraded compared to sample B as both samples show similar characteristics in terms of RMS value, PL intensity and PL FWHM. To obtain a strain-compensated SL using the shutter sequence described in Figure 1-a, the In and Sb shutter opening times could therefore be further adjusted. Besides, this growth method can be used for longwave InAs/GaSb SL as already demonstrated [11] [17]. However, in the context of this work, a p-i-n device structure has been grown using the same shutter sequence as sample B (Figure 1-b) to access the electrical performances.

### 4. Midwave InAs/GaSb SL p-i-n photodiode

The p-i-n device structure is presented in Figure 4 and consists of a 1 µm thick active region made of an undoped 7 ML InAs / 4 ML GaSb SL layer. The XRD spectrum of the ω/2θ scan around the GaSb (004) reflection of sample C is plotted in Figure 5. Again, a nearly strain-compensated SL has been achieved demonstrating the reproducibility of the growth method. The measured period is 3.48 nm which is slightly thinner than sample B resulting in a slight change of the cut-off wavelength equal to 5.15 µm at 77K for sample C as we can see from the PL spectra represented in Figure 6. In the inset of Figure 6, the energy band gap $E_g$, defined as the energy at the maximum PL intensity from which we subtracted the thermal contribution, is plotted as a function of temperature, T. This variation can be described using the well-known Varshni's equation [18] (inset Figure 6). The α parameter (0.3 meV/K) is equal to the value already reported in the literature for the same period composition while the β (325K) parameter which represents the quadratic variation of the band gap at low temperature is 125K lower in our case [15]. Concerning the surface roughness, sample C presents a similar surface morphology to sample B with an RMS value of about 0.20 nm for a scan area of 2 µm x 2 µm.

Sample C was processed into photodiodes using standard photolithography. Cr/Au was deposited as bottom and top contact metal. Mesa photodiodes with a diameter ranging from 90 to 440 µm were realized by wet etching. Photoresist was spun onto the surface and baked to protect the etched surface. The sample was then placed in a cryogenic probe station to perform current - voltage (J-V) and capacitance - voltage (C-V) measurement using a Keysight B1500A Semiconductor Device Analyzer.

The measured J-V characteristics at different temperatures from 77K to 300K are represented in Figure 7-a. The current density at a reverse voltage of -50 mV as a function of the inverse of temperature is plotted in Figure -b along with the variation of the diffusion current ($\propto E_g/kT$ with $E_g$ the experimental band gap from Figure 6 and $k$ the Boltzmann constant) and the generation-recombination current (G-R $\propto E_g/2kT$). Up to 110K, we can observe in Figure 7-a a shift towards positive voltage of the forward characteristics meaning that the photodiode is limited by the photocurrent from the 300K background (in the probe station), i.e. the intrinsic dark-current of the photodiode is lower than this photocurrent. According to Figure 7-b, the device is then limited by the G-R current and starts to be diffusion-limited at a temperature of 136K. In dark conditions, SL p-i-n photodiodes are usually limited by the G-R current at low temperatures [19] [20] [21] [22]. If we assume that the G-R current is the main mechanism limiting the intrinsic performance of sample C at 77K, we can estimate the dark-current level from the G-R variation plotted in Figure 7-b. It is equal to 8.3 x 10$^{-8}$ A/cm$^2$ at -50 mV which is comparable to the state-of-the-art of midwave InAs/GaSb SL p-i-n photodiode grown on GaSb substrate [23] [24].

To complete the electrical characterization, C-V measurements at a frequency of $f = 1$ MHz were performed on diodes with different diameters. A typical C-V characteristic obtained at 77K for a diode diameter of 440 µm is plotted in Figure 8. The residual carrier concentration $N_{res}$ is extracted from the slope close to 0V of the (A/C)$^2$ curve with A the diode area (inset Figure 8) [25]. The $N_{res}$ extracted from the average of (A/C)$^2$ measured for different diode diameters is equal to 9.7 x 10$^{14}$ cm$^{-3}$ which is in good agreement with values reported for the same period composition and thickness [15] [22].

## 5. Conclusion

We investigated on the influence of the shutter sequence during the growth of InAs/GaSb SL on the structural, morphological and optical properties. By growing an intentional InSb layer at both interfaces using the migration enhanced epitaxy technique, the SL layer appeared to be in compressive strain on GaSb. However, by exposing each InAs layer to an incident antimony flux for few seconds before growing the GaSb layer, we obtained a nearly strain-compensated SL ($\Delta a/a \sim 0\%$) with a full width at half maximum of 56 arcsec for the first-order SL satellite peak. The sample surface is relatively smooth with a root mean square value of 0.19 nm on a 2 µm x 2 µm area. Following these results and using the same growth conditions, a p-i-n device structure having a cut-off wavelength of 5.15 µm has been grown and photodiodes fabricated. Electrical performances comparable to the state-of-the-art for midwave InAs/GaSb SL photodiodes on GaSb substrate have been obtained with a dark-current level of 8.3 x 10$^{-8}$ A/cm$^2$ at – 50mV and a residual carrier concentration of 9.7 x 10$^{14}$ cm$^{-3}$ at 77K confirming the good material quality obtained by choosing the appropriate shutter sequence.

### Acknowledgement

This project has received funding from the European Union's Horizon 2020 research and innovation programme under the Marie Sklodowska-Curie grant agreement No 743521 (Project acronym: ASISA). The authors would also like to acknowledge the financial support provided by Sêr Cymru National Research Network in Advanced Engineering and Materials.

# Table captions

**Table 1:** RMS values measured by AFM for different scan area.

# Table

| Scan Area | Sample A RMS (nm) | Sample B RMS (nm) |
|---|---|---|
| 2 µm x 2 µm | 0.20 | 0.19 |
| 4 µm x 4 µm | 0.32 | 0.30 |
| 10 µm x 10 µm | 0.39 | 0.35 |

**Table 1**

**Figure captions**

**Figure 1:** Schematic shutter sequence during the SL growth of sample A and sample B.

**Figure 2:** XRD spectra of sample A (top) and sample B (bottom). The SL layer of sample A is under compressive strain with a lattice mismatch of Δa/a ~ 0.479% while it is nearly strain-compensated for sample B (Δa/a ~ 0%).

**Figure 3:** Photoluminescence spectra measured at a temperature of 77K.

**Figure 4:** InAs/GaSb SL p-i-n device structure grown on GaSb substrate. The active region is made of an undoped 7 ML InAs / 4 ML GaSb SL layer.

**Figure 5:** XRD spectrum of the SL p-i-n device structure (sample C).

**Figure 6:** PL spectra measured at different temperatures (from 77K to 200K). In inset, evolution of the energy band gap with temperature which can be described using the Varshni's equation.

**Figure 7**: a) Current density versus voltage measured at different temperatures. b) Current density measured at -50 mV as a function of the inverse of temperature (symbols). The variation of the diffusion and G-R currents are represented by the dash-dot line (red) and short-dash line

(blue), respectively. The photocurrent level from the 300K background is also plotted (dot line, orange).

**Figure 8:** Typical capacitance-voltage characteristic obtained for diodes with a diameter of 440 µm at 77K. In inset, the corresponding $(A/C)^2$ curve versus voltage (symbols) along with the slope around 0V (line).

**Figures**

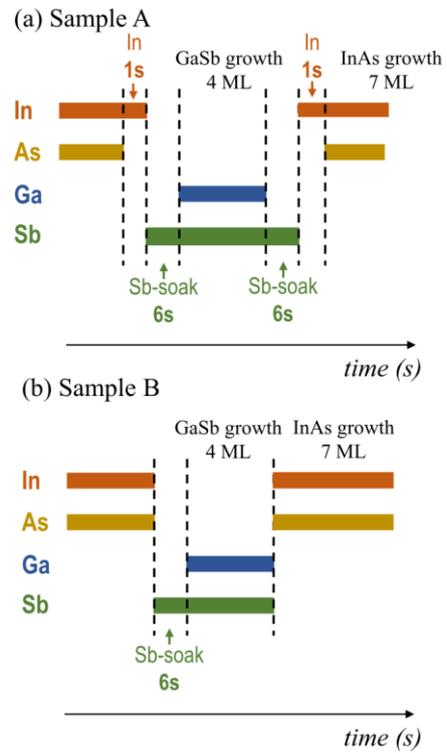

**Figure 1**

M. Delmas *et al.*

Material and device characterization of Type-II InAs/GaSb superlattice infrared detectors

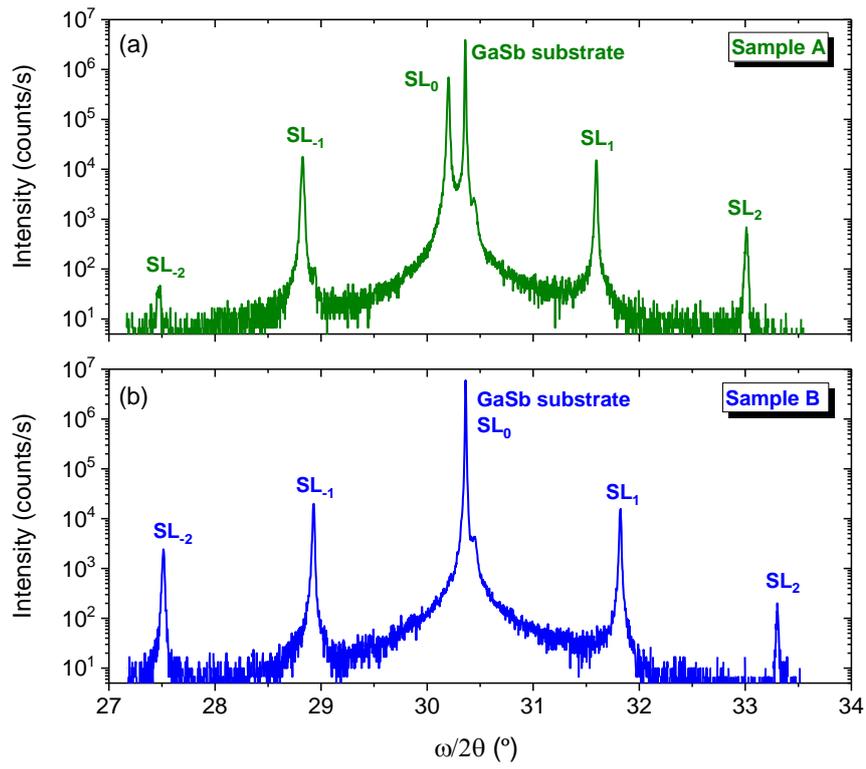

**Figure 2**

M. Delmas *et al.*

Material and device characterization of Type-II InAs/GaSb superlattice infrared detectors

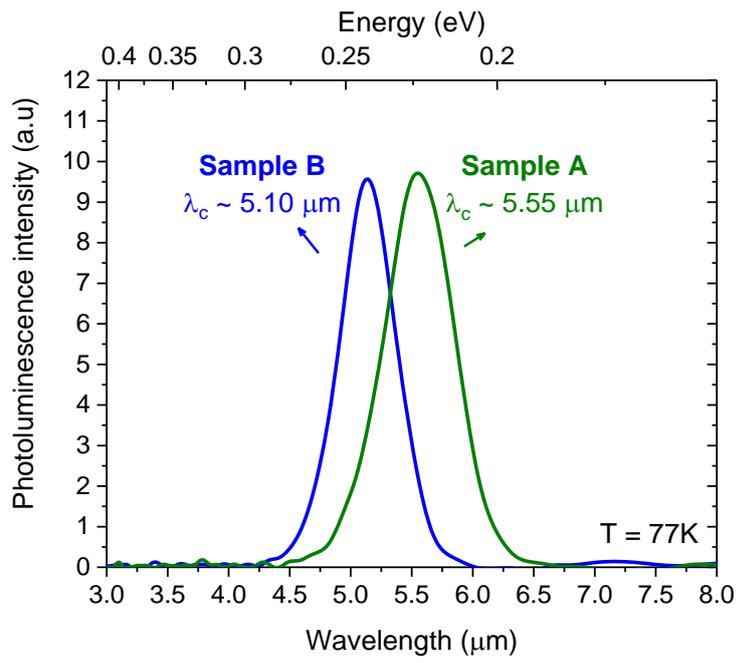

**Figure 3**

M. Delmas *et al.*

Material and device characterization of Type-II InAs/GaSb superlattice infrared detectors

| InAs cap N⁺ ~ 1 x 10¹⁸ cm⁻³ t ~ 20 nm |
| MW T2SL N⁺ ~ 1 x 10¹⁸ cm⁻³ t ~ 60 nm |
| MW T2SL<br>7 ML InAs/4ML GaSb<br>n.i.d<br>t ~ 1 µm |
| MW T2SL P⁺ ~ 1 x 10¹⁸ cm⁻³ t ~ 60 nm |
| GaSb buffer P⁺ ~ 1 x 10¹⁸ cm⁻³ t ~ 100 nm |
| GaSb P-type substrate |

**Figure 4**

M. Delmas *et al.*

Material and device characterization of Type-II InAs/GaSb superlattice infrared detectors

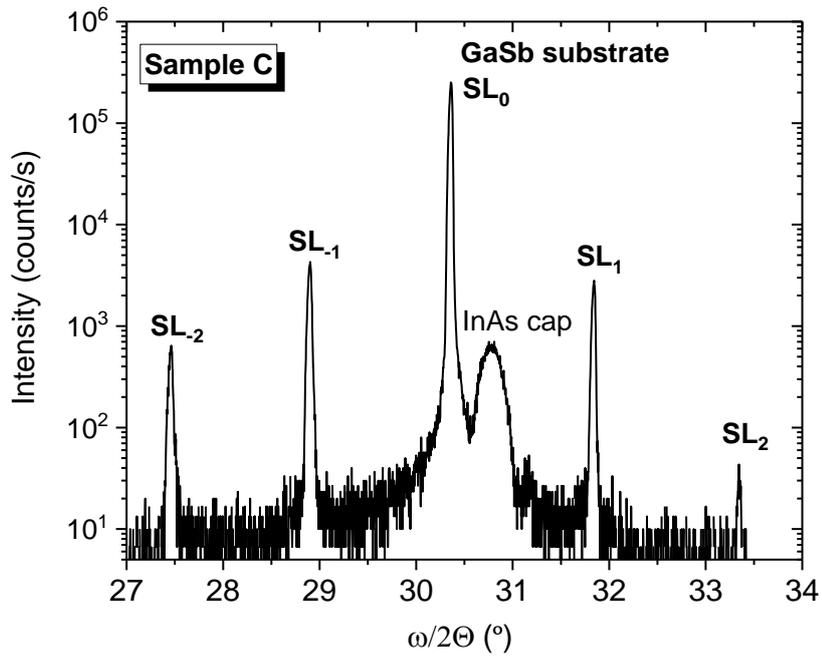

**Figure 5**

M. Delmas *et al.*

Material and device characterization of Type-II InAs/GaSb superlattice infrared detectors

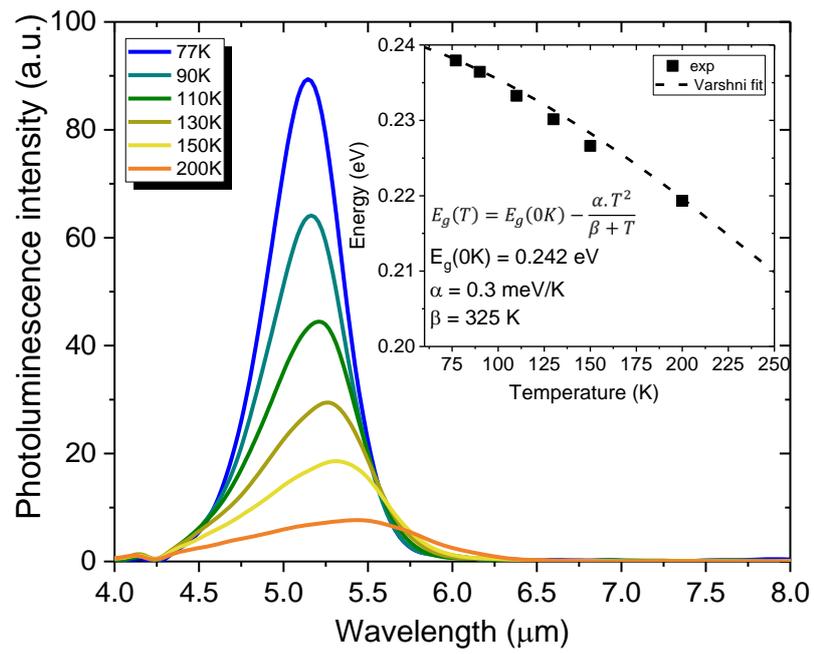

**Figure 6**

M. Delmas *et al.*

Material and device characterization of Type-II InAs/GaSb superlattice infrared detectors

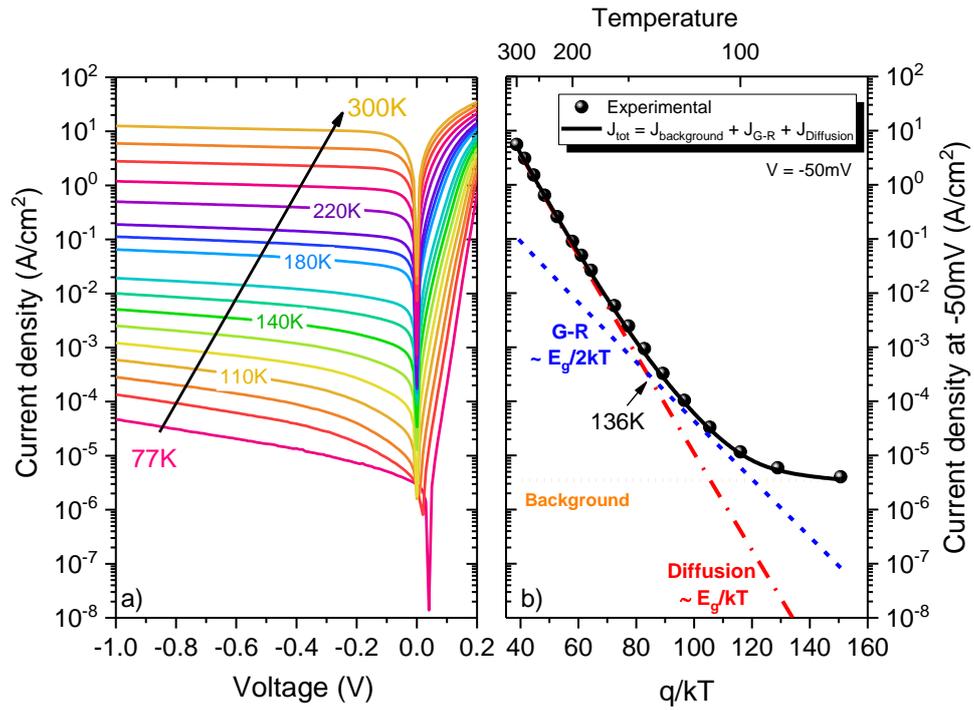

**Figure 7**

M. Delmas *et al.*

Material and device characterization of Type-II InAs/GaSb superlattice infrared detectors

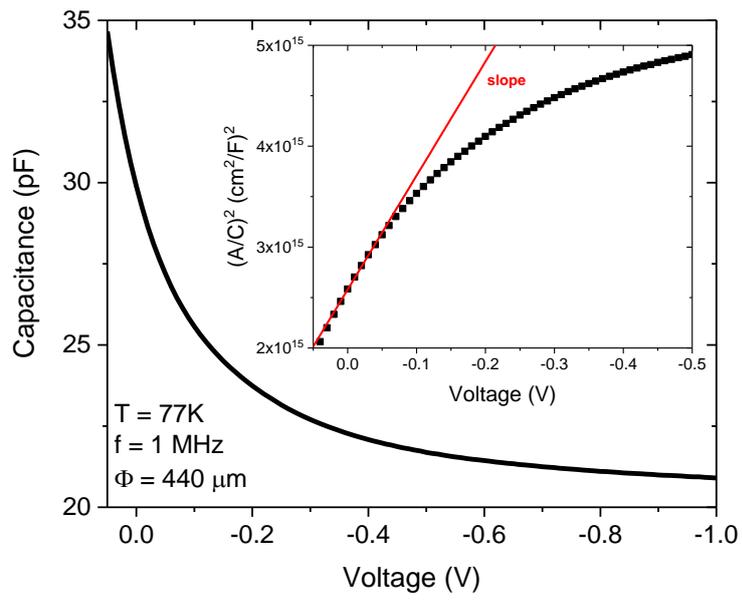

**Figure 8**

M. Delmas *et al.*

Material and device characterization of Type-II InAs/GaSb superlattice infrared detectors